\documentclass[runningheads,anonymous]{llncs}

\usepackage[T1]{fontenc}
\usepackage{graphicx}

\usepackage[table]{xcolor}
\usepackage{tabularx} 
\usepackage{tcolorbox} 
\usepackage{graphicx}
\usepackage{multirow}
\usepackage{rotating}
\usepackage{float}
\usepackage[inkscapelatex=false]{svg}
\usepackage{enumitem}
\usepackage{array}
\usepackage{geometry}

\usepackage{adjustbox} 
\usepackage{booktabs}
\usepackage{url}

\begin{document}
%
\title{Beyond Final Answers:\\Evaluating Large Language Models for Math Tutoring}
\titlerunning{Evaluating Large Language Models for Math Tutoring}

\author{Adit Gupta\inst{1}\orcidID{0000-0002-1218-0630} \and
Jennifer Reddig\inst{2}\orcidID{0009-0000-6731-9401} \and
Tommaso Calò\inst{3}\orcidID{0000-0002-3200-2348} \and
Daniel Weitekamp\inst{2}\orcidID{0000-0003-0079-8000} \and
Christopher J. MacLellan\inst{2}\orcidID{0000-0003-3084-5189}}
\authorrunning{Gupta et al.}

\institute{Drexel University, 3230 Market Street, Philadelphia, PA, USA 
\email{adit.gupta@drexel.edu} \and
Georgia Institute of Technology, North Avenue, Atlanta, GA, USA 
\email{\{jreddig,dweitekamp,cmaclellan\}@gatech.edu} \and
Politecnico di Torino, Torino, Italy 
\email{tommaso.calo@polito.it}}


\maketitle              
\begin{abstract}
Researchers have made notable progress in applying Large Language Models (LLMs) to solve math problems, as demonstrated through efforts like GSM8k, ProofNet, AlphaGeometry, and MathOdyssey.
This progress has sparked interest in their potential use for tutoring students in mathematics.
However, the reliability of LLMs in tutoring contexts---where correctness and instructional quality are crucial---remains underexplored.
Moreover, LLM problem-solving capabilities may not necessarily translate into effective tutoring support for students.
In this work, we present two novel approaches to evaluate the correctness and quality of LLMs in math tutoring contexts.
The first approach uses an intelligent tutoring system for college algebra as a testbed to assess LLM problem-solving capabilities.
We generate benchmark problems using the tutor, prompt a diverse set of LLMs to solve them, and compare the solutions to those generated by the tutor. 
The second approach evaluates LLM as tutors rather than problem solvers.
We employ human evaluators, who act as students seeking tutoring support from each LLM.
We then assess the quality and correctness of the support provided by the LLMs via a qualitative coding process.
We applied these methods to evaluate several ChatGPT models, including 3.5 Turbo, 4, 4o, o1-mini, and o1-preview.
Our findings show that when used as problem solvers, LLMs generate correct final answers for 85.5\% of the college algebra problems tested.
When employed interactively as tutors, 90\% of LLM dialogues show high-quality instructional support; however, many contain errors---only 56.6\% are entirely correct.
We conclude that, despite their potential, LLMs are not yet suitable as intelligent tutors for math without human oversight or additional mechanisms to ensure correctness and quality.

\keywords{Intelligent Tutors, Large Language Models, Generative AI, Math Education}

\end{abstract}

\section{Introduction}
Large Language Models (LLMs) have started to exhibit moderate proficiency at mathematical problem solving. For example, GPT-4 correctly solves over 90\% of the problems in GSM8K benchmark \cite{cobbe2021gsm8k} and approximately 80\% of the problems in the MATH benchmark \cite{fang2024mathodyssey} using advanced prompting techniques \cite{openai2023}. Although these results indicate progress, there are still many limitations. Findings from the GSM-Symbolic benchmark \cite{apple_paper} suggest that LLMs struggle with perturbed or novel problem formulations that are easily solved by humans, indicating that their relatively high performance on standard benchmarks is partially due to memorization. Furthermore, LLM performance remains inconsistent across different problem classes, in contrast to traditional intelligent tutors, which are developed to provide 100\% accurate support. These inconsistencies warrant the need for a deeper investigation into the capabilities, limitations, and implications of LLM for education.

Companies such as Duolingo and Khan Academy have started to leverage LLMs to offer personalized learning experiences, facilitate interactive problem-solving, and provide real-time feedback to learners. However, significant challenges remain to ensure the accuracy, reliability, and adaptability of LLMs in tutoring settings. Despite their remarkable capabilities, studies have shown that LLMs frequently produce plausible yet incorrect solutions to complex mathematical problems, especially in areas that require precise calculations and multi-step reasoning \cite{apple_paper,hendrycks2021math}. In mathematics, not only is the correctness of the final answer crucial, but also the quality of stepwise guidance that fosters effective learning. One recent classroom study comparing LLM-tutoring to traditional classroom instruction showed positive results \cite{kestin2024ai}. However, considering that LLMs likely produce errors in around 10\% of responses---using the best GSM8k performance as an optimistic measure---there is a possibility that they may do more harm than good. The manner in which LLMs confidently ``hallucinate'' incorrect, yet seemly plausible information is a recipe for several possible negative effects \cite{huang2023survey}. In the best case, students' trust in LLM tutors may be rightfully eroded upon recognizing mistakes. In the worst case, LLM hallucinations may lead students to form misconceptions that compromise future learning. 

LLM tutors mark an unusual inflection in the history of intelligent tutoring systems. It has been known for decades that automated computer-delivered tutors produce learning gains comparable to or greater than those of human tutors \cite{vanlehn2011relative,kulik2016effectiveness}, who famously provide learning gains up to two standard deviations higher than those from traditional classroom instruction \cite{bloom1984_2sigma}. The original artificial intelligence (AI) tutors: hard-coded intelligent tutoring systems, have found success by taking a cognitivist approach to tutoring---tracking and quantifying student knowledge by comparison to an expert model \cite{corbett1994knowledge}, and then adapting instruction accordingly \cite{nwana1990intelligent,vanlehn2006behavior}. As compelling as LLMs' generative capabilities are, when used as standalone tutors, they arguably mark a regression in actual AI tutoring capabilities compared to traditional ITSs since they are consistently inaccurate and lack the cognition-oriented adaptivity of prior approaches.

This study evaluates the potential of LLMs in educational contexts by systematically assessing their performance on structured algebra tasks. We selected algebra for this study, given its long-standing use in previous research on intelligent tutoring systems \cite{koedinger1997intelligent,heffernan2014assistments,aleven2006cognitive}. This study aims to investigate the following questions:

\begin{itemize}
    \item \textbf{RQ1}: How accurately can LLMs generate solutions to the kinds of algebra problems currently supported by intelligent tutoring systems?
    \item \textbf{RQ2}: What is the accuracy and quality of the tutoring support provided by LLMs (e.g., scaffolding, hints, and feedback) on these algebra problems?

\end{itemize}

We employ two techniques to explore these questions: (1) an automated approach that uses an existing algebra tutor as a testbed for evaluating LLM problem solving and (2) a qualitative approach to assess the quality and correctness of LLM dialogues generated by having evaluators interactively prompt an LLM for tutor support. 
For the second method, we also conducted a thematic analysis \cite{Braun2006} to identify and categorize observations about LLM tutoring behaviors.

The findings of this study contribute to the field of intelligent tutoring systems by providing empirical evidence on the strengths and limitations of LLMs in math tutoring contexts, thus enriching the ongoing discourse on the role of AI in supporting learning. Specifically, our study makes the following contributions:
\begin{itemize}
    \item We introduce a novel method that uses intelligent tutors as testbeds for evaluating LLM problem solving.
    \item We introduce a second method for interactively evaluating LLM tutoring correctness and quality.
    \item We show that while LLMs largely generate responses aligned with pedagogical best practices, they frequently contain mistakes and inaccuracies, suggesting they are not yet ready for direct in-class deployment.
    \item We offer actionable guidelines for developers, emphasizing how LLMs can support aspects of tutoring, such as question generation and hint production, rather than serving as comprehensive, standalone tutoring solutions.
\end{itemize}

\section{Related Works}
Researchers have begun exploring the use of LLMs in educational applications. The existing literature shows that LLMs can generate worked examples and guide structured problem solving. For example, WorkedGen \cite{workedgen2023} uses prompt chaining and one-shot learning to produce interactive programming examples. Although user studies indicate that 77\% of students found WorkedGen helpful, such self-reported feedback does not necessarily confirm improved learning outcomes. Similarly, Jamplate \cite{jamplate2023} harnesses AI-powered templates for idea generation, providing reflection-based scaffolding, but noting a tendency toward reduced critical thinking among students.

Although these studies highlight the potential of LLMs to create structured examples and facilitate reflective engagement, researchers must develop a consistent, stepwise evaluation framework for algebraic or multi-step reasoning tasks. Existing benchmarks, such as GSM8K \cite{cobbe2021gsm8k}, assess the accuracy of the final answer rather than examining the detailed intermediate steps or the iterative feedback necessary for model-tracing \cite{heffernan2008expanding} in a typical tutoring context. As a result, there is still a need for a more systematic methodology that tests how effectively LLMs handle multi-step problems and adapt to the pedagogical requirements of a tutoring environment.

Another growing area of research investigates the use of LLMs for tutoring in various domains for non-fluent English speakers. For example, a comparative study of models such as GPT-4, Llama-2-ko-DPO-13B, and eT5-chat reveals trade-offs between individualization and correctness \cite{language_tutoring2023}. Smaller models provided more personalized interactions, while GPT-4 exhibited greater correctness but less personalized assistance. Tutoring is an immensely personal activity, and both correctness and individualization are needed. These studies demonstrate the need for more investigation into LLM shortcomings in stepwise instruction, as well as how to better integrate LLM into existing intelligent tutoring platforms. Moreover, current studies often prioritize correctness of the final answer, overlooking the quality of intermediate steps that are crucial for meaningful learning \cite{xia2024reasoning}. For example, in mathematics education, breaking problems down into their steps ensures that students grasp foundational concepts rather than simply arriving at the correct solution. 

The work in this paper aims to fill this gap by introducing a novel method that evaluates LLM performance on a wide range of math questions from college algebra, generated from the Apprentice Tutors platform \cite{gupta2024intelligent}. This platform was designed as a web-based intelligent tutoring platform to support personalized learning in mathematics. The platform supports more than ten tutors including topics like radicals, factoring polynomials, and solving logarithmic equations.

\section{Methodology}
We employ two complementary approaches to evaluate LLMs. First, we developed an automated approach that uses an existing intelligent tutoring system to assess LLM problem-solving accuracy. We generate problems from the tutor and submit them to multiple LLMs. We then use the tutor expert model to evaluate the correctness of the LLM responses. Second, to evaluate LLMs as tutors, we had evaluators interactively engage with the LLMs to request tutoring guidance as if they were students. We then qualitatively evaluated the tutor support generated by the LLMs.

We collected and analyzed data from both approaches to analyze the strengths and limitations of LLMs in structured problem-solving tasks. Figures \ref{fig:benchmark-process} and \ref{fig:student-interaction} illustrate the workflows for these methodologies, which we describe in further detail below.  

\subsection{Evaluating LLM using Intelligent Tutors as Testbeds}

\begin{figure*}[t]
    \centering
    \includegraphics[width=0.8\textwidth, height=0.75\textheight, keepaspectratio]{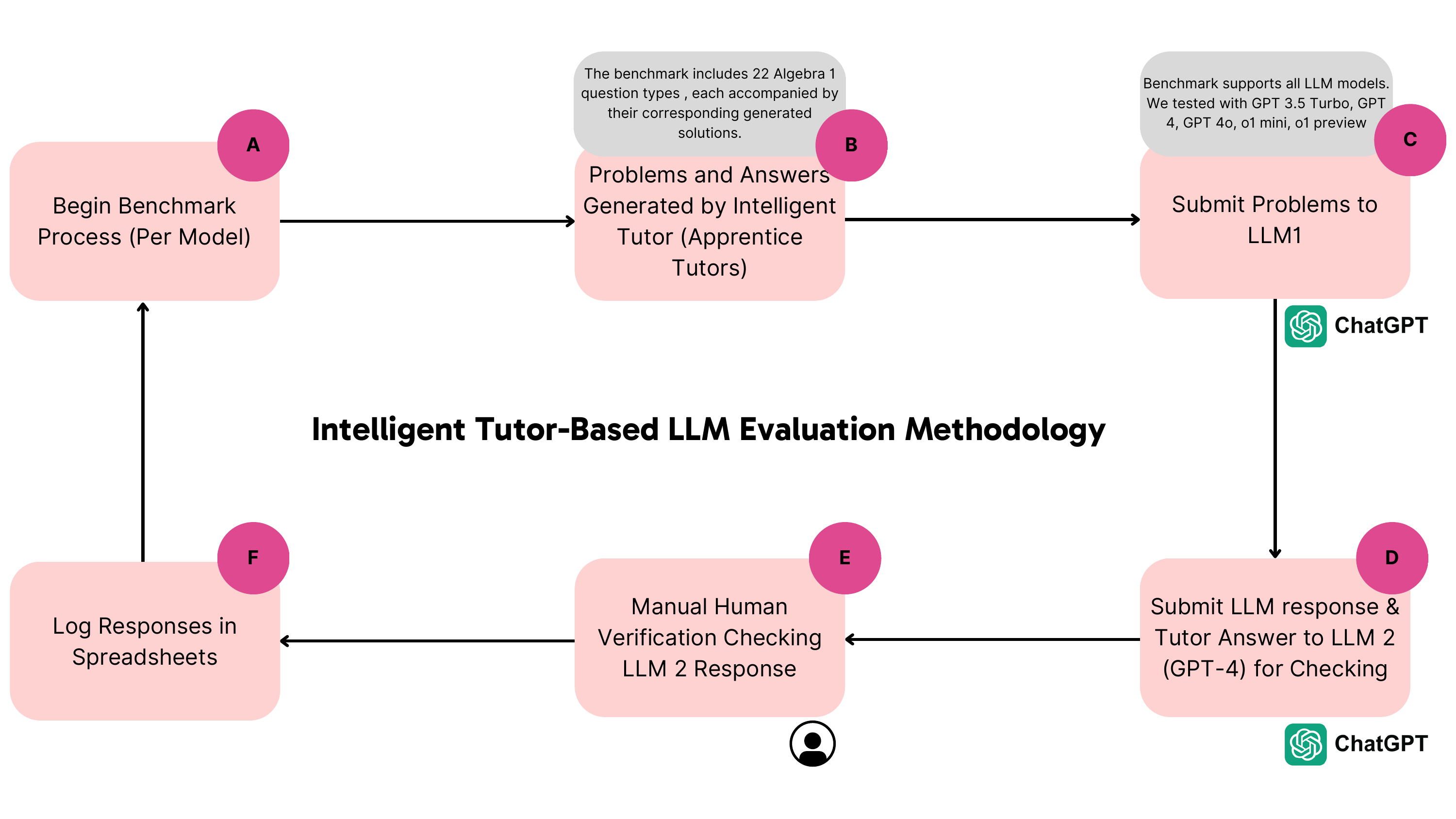} 
    \caption{Our proposed intelligent tutor-based LLM evaluation process. The process begins with the testbed setup - hosted on Google Colab (\textbf{A}), followed by generating problems and solutions using the intelligent tutor (Apprentice Tutors, in our case) (\textbf{B}). For our evaluation, 22 types of problems are then submitted to LLM models such as GPT-3.5 Turbo and GPT-4 (\textbf{C}). Responses from each LLM are checked by submitting them along with the tutor answers to a second LLM (\textbf{D}). We then performed manual human verification to validate the accuracy of the second model's responses. 
    Finally, all results are logged into a performance tracker spreadsheet (\textbf{F}).}
    \label{fig:benchmark-process}
\end{figure*}

We developed our evaluation system in Python to automate tutor problem generation, LLM interaction, and response evaluation. This allowed us to systematically test multiple LLMs on a variety of educational tasks. For this study, we evaluated GPT-3.5 Turbo, GPT-4, GPT-4o, o1 Mini, and o1 Preview. Although these models were the focus of this analysis, the benchmarking tool is designed to be extensible to other tutor content and can be easily adapted to test other models, such as Google's Gemini, Anthropic's Claude series, or Deepseek's open models.\footnote{To maintain anonymity, we have not shared the link to the automated benchmark or data. However, upon acceptance, we will provide the GitHub link to all our code and data.}

The workflow for the tutor-based evaluation process is outlined in Figure \ref{fig:benchmark-process}. The process begins with the testbed setup (\textbf{A}), where we define parameters for the evaluation, including the number and type of algebra problems to be tested. For this study, we identified 22 problem types from the Apprentice Tutors platform and generated five problems of each type. The problems and their corresponding step-by-step solutions were generated directly from the Apprentice Tutors software (\textbf{B}). The generated problems were then submitted to each LLM (\textbf{C}), which was prompted to produce a solution. The exact prompt provided to each LLM was: 

\tcbset{
    colback=white, colframe=black, fonttitle=\bfseries,
    boxrule=0.5mm, arc=0mm, outer arc=0mm,
}

\begin{tcolorbox}[title=Math Problem Solving Prompt]
\small
You will be solving the math questions that are provided as strings. Your task is to parse each question, solve it step-by-step, and provide the final answer in LaTeX format.

Here are the math questions and their answers for verification:
{$<$question\_answer\_text$>$}

Now, here are some new math questions that need answers:
{$<$next\_question\_text$>$}

For each question, think through the $<$problem\_type$>$ problem step-by-step in $\langle$scratchpad$\rangle$ tags. Break down the problem into smaller sub-problems if necessary, and solve each one in a logical order. Show your work and reasoning at each step.

After you have thought through each problem and arrived at a final answer, confirm that it matches the provided answer in LaTeX format inside the corresponding $\langle$answer$\rangle$ tag.
\normalsize
\end{tcolorbox}

The benchmarking system processed the response from the initial LLM prompt and extracted the output, generating a structured list of questions paired with their corresponding answers produced by the LLM. The LLM responses were then evaluated by submitting the original problem and the generated solution to a second LLM (\textbf{D}) to verify accuracy and logical consistency. We used GPT-4 as the evaluation model in all tests. Each LLM was evaluated sequentially and the results were recorded and analyzed before proceeding to the next model. The exact prompt used by the second LLM to evaluate each answer was:

\begin{tcolorbox}[title=LLM Evaluation Prompt]
\small
Just say True or False (nothing else): does  <LLM\_generated\_response> equal the same as  <ground\_truth\_response\_from\_tutor>?
\normalsize
\end{tcolorbox}

To further validate the quality of the LLM evaluation, we performed manual human verification (\textbf{E}). During this process, reviewers compared the ground-truth responses generated by the Apprentice Tutors platform with the responses produced by the LLM and the correctness assessments provided by the second LLM. In certain cases, discrepancies arose due to differences in interpretation, such as when the second LLM marked an expression like \(\sqrt{4}\) as incorrect because it expected the simplified answer of \(2\). These instances were noted and the human reviewer marked the answer as correct if it was mathematically accurate in its final form. However, stepwise solutions were also considered, ensuring that intermediate simplifications (e.g., distinguishing between $\sqrt{4}$ and 2 when necessary) aligned with expected problem-solving conventions.

Finally, all data collected, including LLM responses, human evaluations, and any discrepancies identified during the validation process, were systematically recorded in a performance tracker spreadsheet (\textbf{F}). This structured logging approach facilitated detailed analysis and allowed for a robust comparison between different LLM models and problem types. The goal was to gain insights into their performance and limitations. 

\subsection{Evaluating LLMs via Interactive Prompting}
We conducted a second study to assess how the LLMs perform when interacting with learners as tutors (see Figure \ref{fig:student-interaction}). This study was designed to provide a qualitative perspective on the educational capabilities of the model, in contrast to the previous automated evaluation of problem-solving capabilities. We performed manual evaluation of the LLMs, using a standardized variation of the prompt from Salman Khan's widely cited ChatGPT interview \cite{khan2023chatgpt} to ensure consistency between sessions. By comparing these human-guided interactions with the outputs of the intelligent tutor, we investigated how well the step-by-step guidance of LLMs align with real user queries and misconceptions in a math tutoring context. Here is the prompt that was used:

\begin{figure*}[t]
    \centering
    \includegraphics[width=0.75\textwidth, height=0.75\textheight, keepaspectratio]{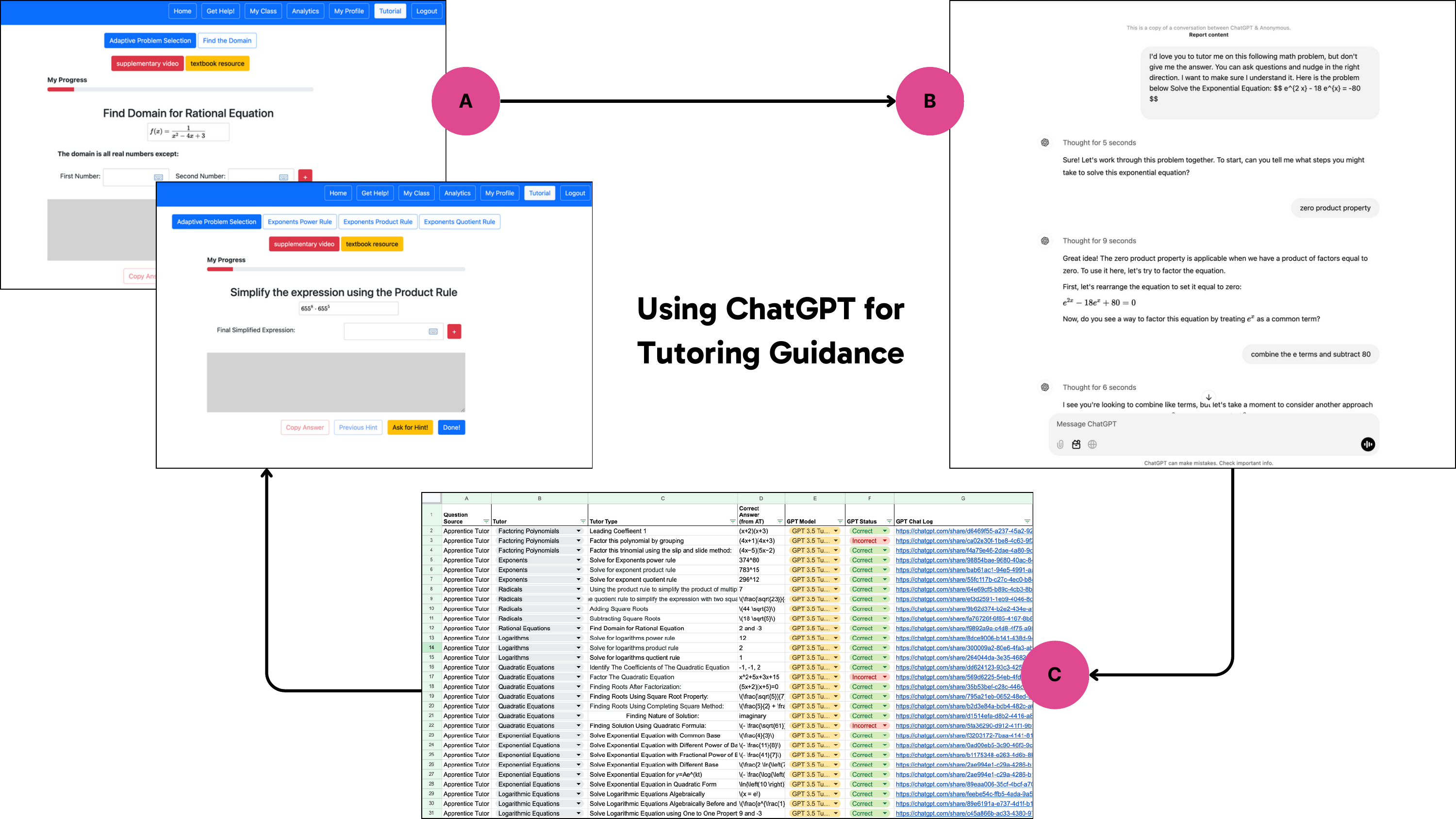}
    \caption{Our proposed process for evaluating an LLM via interactive prompting. This diagram illustrates the process using ChatGPT's chat interface. The workflow begins by having evaluators prompt ChatGPT to provide tutor guidance on tutor problems as if they were students (\textbf{A}). Evaluators interactively submit queries and receive step-by-step guidance from ChatGPT (\textbf{B}). The evaluators systematically log each chat dialogue in a spreadsheet tracker for analysis (\textbf{C}).}
    \label{fig:student-interaction}
\end{figure*}

\begin{tcolorbox}[title=Interactive Tutoring Prompt]
\small

I'd  love you to tutor me on this following math problem, but don't give me the answer. 
You can ask questions and nudge me in the right direction. I want to make sure I understand it. 
Here is the problem below. $<$Problem$>$

\normalsize
\end{tcolorbox}

The problems were taken directly from the Apprentice Tutors and entered into ChatGPT. The evaluators interacted with the model as it guided them through problem-solving, responding to ChatGPT’s hints and prompts as they progressed. After each session, they logged a link to the chat history, the final answers provided by ChatGPT, and whether the responses matched the correct answers generated by the Apprentice Tutors. An example of this recording is shown in part B of Figure \ref{fig:student-interaction}. 

After collecting all responses, two independent reviewers assessed the interactions to answer the following questions about each tutoring dialogue:
\begin{itemize}
    \item \textbf{Quality:} Do the steps represent a high-quality tutoring interaction?
    \item \textbf{Correctness:} Were all the LLM responses in the dialogue correct?
\end{itemize}

\begin{table*}[b!]
\centering
\renewcommand{\arraystretch}{1.5} 
\setlength{\tabcolsep}{8pt}       
\scriptsize
\begin{tabularx}{\textwidth}{|X|X|X|X|X|}
\hline
& \multicolumn{2}{c|}{\textbf{Low}} & \multicolumn{2}{c|}{\textbf{High}} \\
\hline
\textbf{Criterion} & \textbf{1 (Poor)} & \textbf{2 (Fair)} & \textbf{3 (Good)} & \textbf{4 (Excellent)} \\
\hline
Clarity of Explanation &
Explanations are frequently unclear, leading to student frustration. &
Explanations are sometimes unclear or overly complex. &
Explanations are generally clear, with occasional confusion. &
Explanations are consistently clear, logical, and easy to understand. \\
\hline
Feedback &
Feedback is rare or unhelpful, lacking specificity. &
Feedback is provided, but is sometimes vague or untimely. &
Feedback is generally constructive, specific, and timely. &
Feedback is consistently constructive, specific, timely, and enhances student understanding. \\
\hline
Scaffolded Support &
The tutor provides too little or too much support, hindering learning. &
Support is inconsistent, with occasional mismatches to the student's needs. &
Support is generally well-matched to the student’s needs, with appropriate scaffolding. &
Support is perfectly calibrated, with scaffolding gradually reduced as the student gains confidence. \\
\hline
Problem-Solving Strategies &
Problem-solving strategies are not discussed or modeled. &
Some strategies are mentioned, but with limited discussion or modeling. &
Effective problem-solving strategies are discussed and modeled. &
The tutor effectively models and teaches problem-solving strategies, encouraging independent thinking. \\
\hline
Encouragement and Reinforcement &
The tutor provides little to no encouragement, leading to a negative learning atmosphere. &
Encouragement is sporadic, sometimes failing to motivate the student. &
The tutor provides positive reinforcement, generally maintaining a supportive atmosphere. &
The tutor consistently encourages and motivates the student, fostering a highly positive learning environment. \\
\hline
\end{tabularx}
\caption{Rubric for evaluating the quality of interactive tutoring dialogues in terms of five criteria.}
\label{tab:tutoring_interaction}
\end{table*}

To answer the first question and classify response {\bf quality}, each dialogue was evaluated with respect to the structured rubric shown in Table \ref{tab:tutoring_interaction}. Each criterion was scored on a scale from 1 to 4. The rubric was designed to measure several key aspects of tutoring quality. Specifically, it evaluated (A) the correctness of explanations, (B) the depth of scaffolding, and (C) the alignment with instructional best practices. This structured approach aimed to reflect established learning science principles. We summed the scores across all criteria. If the total score was 10 or below, the response was categorized as ``No'' (not good quality); if the score was above 10, it was categorized as ``Yes'' (good quality). Once the scores were converted into Yes/No label, we measured inter-rater reliability using Cohen’s Kappa \cite{McHugh2012} to assess agreement between reviewers and confirm the robustness of our classifications.

To answer the second question and assess response {\bf correctness}, each reviewer also independently assessed whether all the LLM generated content was correct. These evaluations consisted of considering each LLM response from the dialogue, and noting any mistakes or errors. If there were any errors, then the dialogue was coded as incorrect, otherwise it was recorded as correct. Similar to question 1, we measured inter-rater reliability of the two evaluations using Cohen’s Kappa. We also conducted a thematic analysis \cite{Braun2006} of the LLM responses to identifying recurring patterns in tutoring interactions. The goal was to identify and categorize common patterns, counting their frequency and noting whether they corresponded to positive or negative tutoring behaviors.

\section{Results and Analysis}
We present the results of the two evaluation methods used to assess the performance of LLMs in math tutoring contexts. We first present the results from our tutor-based evaluation method and then the results from evaluators interactively prompting the LLMs as if they were students.

\subsection{Tutor-Based LLM Evaluation Results}

Table \ref{tab:api_based_results} summarizes the results from the tutor-based evaluation. The apprentice tutors had 22 problem types and we sampled 5 problems of each type to produce a test set that contained 110 problem and answer pairs.

\begin{table}[t]
\centering
\small 
\caption{The tutor-based LLM evaluation results, which only assessed the final answer.}
\begin{tabularx}{\columnwidth}{l>{\centering\arraybackslash}X>{\centering\arraybackslash}X>{\centering\arraybackslash}X>{\centering\arraybackslash}X}
\toprule
\textbf{Model} & \textbf{\# Problem Types} & \textbf{\# Problems} & \textbf{\# Correct} & \textbf{Accuracy} \\
\midrule
GPT-3.5 Turbo & 22 & 110 & 85 & 77.3\% \\ 
GPT-4         & 22 & 110 & 83 & 74.5\% \\ 
GPT-4o        & 22 & 110 & 107 & 97.3\% \\ 
o1-mini       & 22 & 110 & 101 & 91.8\% \\ 
o1-preview    & 22 & 110 & 94 & 85.5\% \\ 
\midrule
    \textbf{Overall Avg.} & 22 & 110 & 94 & 85.5\% \\
    \bottomrule
\end{tabularx}
\label{tab:api_based_results}
\end{table}

We identified twenty-five instances (6.3\% of total responses) where the second LLM marked answers incorrectly.
From our observations, the second LLM would incorrectly mark the answer when comparing the tutor-generated response to the LLM-generated response for the following reasons: \textbf{(1)} a mismatch in the operational order (e.g., \((3 + 6) \times 2\) vs. \(3 + (6 \times 2)\)), \textbf{(2)} differences in simplification (e.g., \(\frac{2}{4}\) vs. 0.5), and \textbf{(3)} differences in operator context (e.g., multiplication represented by ``x'' vs. ``*'').

\subsection{Interactive Prompting-Based LLM Evaluation}

The evaluators prompted each of the five models to provide tutoring support on the same set of 30 problems, resulting in a total of 150 LLM dialogues. The two reviewers independently analyzed each dialogue to classify whether it was high quality (using rubric from Table \ref{tab:tutoring_interaction}) and fully correct. We also evaluated the final answer accuracy from each dialogue by comparing it to the tutor solution.
Table \ref{tab:manual_results} shows the results of these assessments, breaking out the accuracy of final LLM answers alongside reviewer assessments of the quality and correctness of the LLM dialogues.
We calculated Cohen’s Kappa ($\kappa$) to evaluate the reviewer agreement for both the quality and correctness ratings. This score, which ranges from 0 to 1, represents agreement after correcting for chance. A score greater than 0.7 is typically viewed as strong agreement. For the quality ratings, Cohen’s $\kappa_{Quality} \approx 0.85$, and for the assessment of whether the LLM dialogues were entirely correct, Cohen’s $ \kappa_{Correctness} \approx 0.82$. These scores indicate very strong agreement between the independent reviewers.

\begin{table}[t]
    \centering
    \renewcommand{\arraystretch}{1.1} 
    \setlength{\tabcolsep}{4pt} 
    \caption{Assessments of LLM tutoring interaction quality and accuracy. The columns show final answer accuracy as well as the percentage of LLM dialogues that were classified as high quality and fully correct, as indicated by reviewers R1 and R2. The number of problems in each case is noted in parentheses.}
    \begin{tabular}{l >{\centering\arraybackslash}p{2cm} >{\centering\arraybackslash}p{3cm} 
                >{\centering\arraybackslash}p{1.6cm} >{\centering\arraybackslash}p{1.6cm} 
                >{\centering\arraybackslash}p{1.6cm} >{\centering\arraybackslash}p{1.6cm}}
        \toprule
        \textbf{Model} & \textbf{\# Problems} & \textbf{Final Accuracy} & \multicolumn{2}{c}{\textbf{\% High Quality}} & \multicolumn{2}{c}{\textbf{\% Fully Correct}} \\
        \cmidrule(lr){4-5} \cmidrule(lr){6-7}
        & & & R1 & R2 & R1 & R2 \\
        \midrule
        GPT 3.5 Turbo & 30 & 90.0\% (27) & 90.0\% (27) & 90.0\% (27) & 46.7\% (14) & 53.3\% (16) \\
        GPT 4         & 30 & 83.3\% (25) & 93.3\% (28) & 93.3\% (28) & 43.3\% (13) & 50.0\% (15) \\
        GPT 4o        & 30 & 93.3\% (28) & 90.0\% (27) & 90.0\% (27) & 70.0\% (21) & 80.0\% (24) \\
        o1 mini      & 30 & 86.7\% (26) & 86.7\% (26) & 80.0\% (24) & 56.7\% (17) & 43.3\% (13) \\
        o1 preview   & 30 & 90.0\% (27) & 90.0\% (27) & 96.7\% (29) & 73.3\% (22) & 50.0\% (15) \\
        \midrule
        Overall Avg. & 30 & 88.6\% & \multicolumn{2}{c}{90.0\%} & \multicolumn{2}{c}{56.6\%} \\
        \bottomrule
    \end{tabular}
    \label{tab:manual_results}
\end{table}

\begin{table}[htbp]
\centering
\small
\caption{Summary of key observations from the interactive tutoring evaluation.}
\label{tab:key_observations}
\begin{tabularx}{\linewidth}{>{\raggedright\arraybackslash}X c c}
\toprule
\textbf{Observation} & \textbf{Occurrences} & \textbf{Sentiment} \\
\midrule
The final answer was correct, even though there were incurabilities in the sub-steps. & 6 & Negative \\
Even though the prompt was not to share an answer, it was possible to obtain the answer by manipulating responses (via yes/no questions). & 4 & Negative \\
For topics like factoring, there was an overemphasis on teaching basics (e.g., multiplying) instead of demonstrating specific methods (e.g., ``slip and slide''). & 4 & Negative \\
LLM over-indexes on ensuring the final answer is correct rather than emphasizing the student’s step-by-step skill acquisition. & 3 & Negative \\
LLM occasionally produces an incorrect conclusion and refuses to accept a correct student answer. & 4 & Negative \\
Sub-steps are sometimes flagged as incorrect even though they are actually correct. & 3 & Negative \\
Difficult math notation (e.g., quadratic expressions) can be challenging to input from a standard keyboard. & 2 & Negative \\
LLM is flexible about answer formats, accepting multiple notational styles. & 3 & Positive \\
LLM excels at generating hints and extra worked examples to support instruction. & 2 & Positive \\
LLM provides encouraging feedback and positive reinforcement, which could benefit student well-being. & 7 & Positive \\
LLM nudges students to answer queries in sequence when they attempt to skip ahead. & 2 & Positive \\
\bottomrule
\end{tabularx}
\end{table}

Finally, the reviewers evaluated each model’s performance, documenting key behavioral patterns and noting any common issues. Table \ref{tab:key_observations} summarizes these findings, classifying observations as positive or negative based on their potential impact on learners. This analysis highlights the strengths and weaknesses of each model’s tutoring approach, offering insight into their effectiveness in guiding students through problem-solving.

\section{Discussion}
Both of our evaluation methods suggest that the LLMs show reasonable final answer accuracy. Our tutor-based evaluation showed that GPT 4o had the highest final-answer accuracy at 97.3\%. In the interactive prompting-based evaluation, we found that GPT-4o also got the highest accuracy at 93.3\%.
Although these accuracies seem reasonable, it still means that these models will generate incorrect final answers for about 1 in 18 problems. We were also surprised to find that GPT-4 performed the worst and that model rankings changed based on the evaluation.

Newer models often performed worse than older ones, despite expected improvements. This suggests that LLM performance can no longer be expected to improve with each new model release and that there are continued gaps in how LLMs process multi-step math questions.
In terms of final answer accuracy, we also observed that the interactive prompt-based evaluation results were higher than those from the tutor-based evaluation, probably because human testers engaged in multi-turn interactions, allowing LLMs to refine responses. In contrast, the tutor-based evaluation provided only a single prompt, requiring the model to solve problems correctly in one step. 

During our interactive prompting-based evaluation, we found that LLMs generate high-quality tutoring support most of the time.
Although GPT-4 had the lowest final answer accuracy, it scored near the top in terms of quality, with 93.3\% of its chat dialogues being classified as having high pedagogical quality. 
Although the LLMs achieved reasonable final answer accuracies, we found that when considering their entire tutoring dialogues they often would make mistakes. Along this dimension, GPT-4o achieved the best performance, with 75\% of its dialogues being classified as fully correct (averaging across the two reviewers). 
Across all five of the LLMs only 56.6\% of the tutor dialogues were entirely correct.
This suggests that nearly 1 in 2 interactive LLM tutoring sessions with a student will likely contain errors.
These results suggest that LLMs error rates are likely much higher would be suggested by benchmarks that only evaluate them in terms of the final answer. This raises concerns about their use as standalone tutors, as less-than-perfect accuracy can harm learners. If one in two dialogues contains errors, students may lose trust in the tutor and, worse, develop misconceptions that hinder future learning.
Our results also suggest that future evaluations must consider the correctness of the entire tutoring dialogue, not just the final answer accuracy. 

Unlike intelligent tutors, which are often developed through meticulous cognitive task analysis~\cite{lovett1998cognitive} to ensure that sub-steps are carefully designed for effective learning, LLMs tend to prioritize arriving at the final answer rather than ensuring students understand the intermediate steps. For example, in factoring problems, the LLMs frequently provided overly general guidance, such as basic multiplication rules, instead of focusing on specific strategies such as the ``slip-and-slide'' method explicitly requested in the task. Although the question will be marked right in the evaluation, not using the specified method reduces the overall quality of the tutoring guidance.

Table \ref{tab:key_observations} summarizes reviewer observations of tutoring interactions. Reviewers noted that the LLMs sometimes refused to accept correct answers, miscalculated sub-steps, or overemphasized fundamentals at the expense of specialized techniques. However, the LLMs also provided flexible response formats, detailed hints, and encouraging feedback. Manual review of chat logs revealed inconsistencies in how LLMs handled intermediate steps. While they often produced correct final answers, sub-steps were occasionally miscalculated or erroneously flagged as incorrect (8 instances, as shown in Table \ref{tab:key_observations}). These errors fell into three categories: (1) simplified vs. unsimplified answers (e.g., \(2\) vs. \(\sqrt{4}\)), (2) differences in term order (e.g., \(\sqrt{3} + \sqrt{4}\) vs. \(\sqrt{4} + \sqrt{3}\)), and (3) formatting mismatches (e.g., missing required LaTeX tags). These issues highlight inconsistencies in how LLMs evaluate semantic equivalence.

Though LLMs may be insufficient compared to ITS when tutoring, recent work demonstrates how LLMs could support ITS in hint generation. By leveraging the expert model of ITS, LLMs can generate correctness feedback personalized to student responses without needing the LLM to perform any mathematical calculations \cite{Reddig2024}.
If integrated with other educational technologies, they could offer several potential benefits. Their ability to generate hints, provide alternative explanations, and accommodate various answer formats makes them flexible and adaptive. However, our finding that LLMs sometimes mark correct answers as incorrect---even when provided with the solutions---suggest that LLMs integrations will need to be carefully evaluated before deployment. 

Additionally, their use of positive reinforcement, such as motivational nudges and encouraging feedback, could fosters an engaging learning environment, as long as all learners receive equal encouragement. For example, several chat logs included statements like, ``Way to go, you are close to the answer!'' or ``That's not right, but let's keep trying.'' These reinforcements might help motivate learners to persist and promote sustained engagement with educational tools. This approach aligns with adult learners' need for constructive feedback and encouragement \cite{gupta2024intelligent}. Future research should systematically evaluate the motivational potential of LLMs interactions and whether they translate into improved learning outcomes. 

\section{Limitations and Future Work}
One limitation of our LLM evaluation methods is that they did not directly evaluate against actual students. We chose our approach because we knew that LLMs have reliability issues and we did not want to cause harm to students by giving them incorrect tutoring guidance during our experiments. Although our approach provides a means of safe, controlled evaluation, it may not fully capture the unique ways in which real students would engage with LLM tutors. A future iteration of this work could involve deploying the system in real-world educational settings and analyzing chat logs generated from authentic student interactions to gain a more comprehensive understanding of LLM performance. However, our work suggests that LLMs make mistakes in just over half of their student tutoring dialogues, so future research must account for the risks to students that this poses.

Furthermore, this study was conducted using an earlier version of the Apprentice Tutors platform, which focused exclusively on math-related questions. The Apprentice Tutors platform has since been expanded to include other types of questions, such as those related to nursing education. Future research could explore how LLMs, including ChatGPT, perform in this and other domains, extending the scope of evaluation to understand their domain-specific adaptability and effectiveness.

Finally, another limitation is that the analysis presented was restricted to a set of LLMs within the ChatGPT family. With the rapid development of new LLMs, such as Google's Gemini, Anthropic's Claude and several open-source LLMs, there is an opportunity to expand this study to evaluate these emerging models too. Comparing performance across a broader range of LLMs would provide a more holistic view of their strengths and weaknesses in educational contexts. Lastly, this study used the publicly available versions of the ChatGPT models accessible online. In practice, commercial production environments often deploy models that are fine-tuned to specific domains or tasks. Evaluating a custom-tuned LLM tailored to specific educational needs could offer a more accurate representation of how these tools would perform in real-world applications. 

\section{Conclusion}
In this study, we evaluated the ability of various LLMs to solve college algebra problems and to interactively provide step-by-step tutor guidance. We evaluated multiple models, including GPT-3.5 Turbo, GPT-4, GPT-4o, o1 Mini and o1 Preview, identifying both their strengths and limitations. The performance results presented in this study, though commendable, are significantly lower than the 100\% accuracy achieved by traditional intelligent tutors on the same set of problems. While we saw an overall final accuracy of 85.5\% with the automated tutor-based evaluation and 88.6\% with our interactive prompting-based evaluation, our analysis of the entire LLM tutoring dialogues showed that only 56.6\% were entirely correct. This discrepancy suggests a core limitation of using LLMs as tutors: while they often generate correct final answers, ensuring the pedagogical soundness and accuracy of intermediate steps remains challenging.

Despite these limitations, LLMs exhibit several capabilities that have the potential to improve learning outcomes. Their flexibility in accepting diverse answer formats, the ability to generate hints and alternative problem explanations, and the use of positive reinforcement, such as motivational nudges, could help foster a more supportive and engaging learning environment. However, there are risks associated with the deployment of LLMs in educational settings. For example, biases within the models may lead to inflexibility in pedagogical approaches, such as internal biases that favor some methods of solving problems over others. Furthermore, inaccuracies in responses---with around one in two dialogues containing errors---can undermine the trust of students in the guidance of the tutor and reduce their confidence in the system. To address these challenges, future work might explore how to leverage their independent capabilities, such as problem generation, hint generation, and positive reinforcement. By balancing these strengths with strategies to manage and mitigate their limitations, LLMs could effectively supplement other educational technologies, such as intelligent tutoring systems.

\section*{Acknowledgments}

This work was generously funded in part from several sources, including the NSF National AI Institutes program (\#2247790 and \#2112532). The views, opinions and/or findings expressed are those of the author and should not be interpreted as representing the official views or policies of the Department of Defense or the U.S. Government. We also thank the members of the Teachable Artificial Intelligence (TAIL) Lab for their feedback and suggestions.

\bibliographystyle{abbrv}
\bibliography{AIED-Paper} 

\end{document}